\documentclass[a4paper]{article}
\usepackage{amsfonts,amssymb,amsthm}
\usepackage[totalwidth=17cm,totalheight=24cm]{geometry}
\usepackage{epsfig}

\newcommand{\R}{{\mathbb R}}

\begin{document}

\title{Errata and Addendum to \\ ``$\Lambda<0$ Quantum Gravity in 2+1 Dimensions''}
\author{Kirill Krasnov \thanks{School of Mathematical Sciences, University of Nottingham,
Nottingham, NG7 2RD, UK}}

\maketitle

\begin{abstract} We correct some errors in the two papers
published with the above title in Class. Quant. Grav. {\bf 19} (2002).
In particular, the correct prescription for computing the probabilities
is given, in that appropriate normalization factors are introduced.
The resulting computation of the semi-classical limit of probabilities 
actually becomes much simpler, and no CFT analysis is necessary.  
In spite of some mistakes, the conclusions of these two papers are to a large 
extent unchanged. In particular, we still get an exponentially small answer 
$\exp{- \beta M}$ for the black hole creation-evaporation probability.
\end{abstract}

\maketitle

Let us first emphasise that the prescription given in the two
papers \cite{Papers} should be viewed as more of an outline of a
theory of $\Lambda<0$ quantum 3d gravity than an actual proposal 
for such a theory. The arguments of \cite{Papers} do lead to quantitative predictions
only in the semi-classical limit of the large radius of curvature,
when it does not matter which exactly CFT is used.

In these note we correct the prescription given in \cite{Papers}, and
correct the answer for the black hole out of two point particles
creation probability. The result is still exponentially small
$\exp{-\beta M}$, where $\beta, M$ are the inverse temperature and mass
of the black hole being created. The interpretation of this result is 
basically unchanged from the one given in \cite{Papers}.

There many gaps in the arguments of \cite{Papers}, some of
which have been filled after the papers were written, some are still to
be settled. The main open issue is still to find the CFT which
is relevant for $\Lambda<0$ quantum gravity and which is referred to
in these papers as a relative of Liouville theory. Once this CFT is found,
the prescription for computing amplitudes formulated in these two papers
can be used for computing probabilities of physically interesting processes.
Today, four years after the papers have been written, the outlined programme
for the construction of the theory of $\Lambda<0$ quantum gravity is
closer to its completion. Important advances have been made in
understanding the Chern-Simons theory of the Lorentz group, see
\cite{Roche} and references therein. Some important
progress has also been achieved in understanding the classical
theory. Thus, the analytic continuation procedure on which the 
papers \cite{Papers} are based was made much more precise 
in the works of Benedetti and Bonsante, see \cite{Ben} and references
therein. More recently, the work \cite{KS} in particular analysed the phase 
space of $\Lambda<0$ gravity, and it was shown that the reduced phase space is the
cotangent bundle over the Teichmuller space of the spatial slice.
The same work also proposed yet another well defined analytic
continuation procedure between the Lorentzian and Euclidean
signature negative curvature 3-manifolds.

\bigskip
We now turn to corrections for the first paper.
\begin{itemize}
\item In several places in the paper (e.g. Page 3982, the paragraph before last on this page, Page 3985, 
last paragraph of Section 3) it is incorrectly stated that manifolds containing 
lines of conical singularities (point particles) can be obtained
by considering discrete groups, subgroups of ${\rm PSL}(2,\R)$, 
containing elliptic elements. This is
in general not true. A group generated, in particular, by elliptic generators,
in general fails to be discrete, and the quotient $H_3/\Gamma$ is not
a manifold. This happens already in the case of a single elliptic generator,
unless the deficit angle is rational. Thus, in general, manifolds containing
point particles are not quotients of $H_3$ and have to be described and
treated differently. A good construction for such manifolds is to
consider a hyperbolic manifold, such as e.g. $H_3$, with a two geodesic
half-planes identified, so that a wedge is cut out. The axis of the transformation
that maps one half-plane into the other becomes particle's worldline. To summarize,
the spaces considered in this paper do exist, they are just not given by 
quotients anymore. All other arguments of the paper go through unchanged. The description of
spaces with particles is more involved than that of
non-singular spaces, and is subject of current interest of the
mathematical community. We refer the reader e.g. to \cite{KS},
where an extensive treatment of all such issues is given.
\item The argument given on Page 3986 explaining why it is
natural for the quantum states to be functionals on the
Teichmuller space is essentially correct, in that it uses the
fact that the reduced phase space of $\Lambda<0$ gravity 
is the cotangent bundle over the Teichmuller space. This
last fact, however, can be shown much more directly,
see \cite{KS} for the proof.
\item The prescription for extracting probabilities that is given in 
the paper needs to be modified, as the numbers one gets using
the prescription of the paper are not normalised. The modification
is straightforward. Let us first consider the case of emission
of a particle by the BTZ black hole. The state (amplitude) for
a black hole with a particle emitted is correctly given by (5.1).
However, to extract the probability, is not enough to
simply square the amplitude. The correct procedure is as
follows. Let us introduce a state $\Psi_{BH}$ describing the BTZ black hole
with no extra point particle. This state is given by the CFT
partition function on the cylinder. This is a state in the
tensor product of two Hilbert spaces, one Hilbert space ${\cal H}$ for each
asymptotic region. The Hilbert space in question is just that of
the CFT. Namely, it is the direct sum ${\cal H}=\oplus V^\Delta$ of Verma modules
$V^\Delta$, where $\Delta$ runs over the set of conformal dimensions
of primary fields of the theory. Thus, the amplitude (state) of the BTZ black hole,
as well as the state of the black hole with a point particle emitted, 
is in ${\cal H}^{\otimes 2}$. To find the probability of emission we need
to find an overlap of these two states. It is easiest to do this by
introducing density matrices that refer only to one of the asymptotic regions,
and then applying the usual rules of quantum mechanics to these density matrices. Thus,
for the black hole, we introduce the density matrix $\rho_{BH}$ by
taking the product of $\Psi_{BH}$ with $\overline{\Psi_{BH}}$ and
summing over the degrees of freedom of one of the asymptotic regions.
As we have sketched in the paper, such a sum is really a sum over
the boundary conditions at that asymptotic region, and its
effect is to glue two Riemann surfaces together, to produce
a ``longer'' cylinder. Both ends of this cylinder now correspond
to the same asymptotic region:
\begin{equation}\label{density-BH}
\rho_{BH}=\sum_b Z_{\rm CFT}\left[\,\,
\lower0.15in\hbox{\epsfig{figure=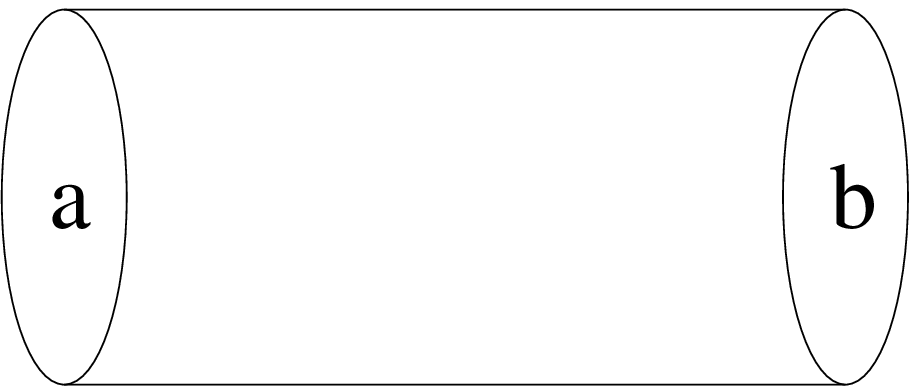,height=0.3in}}
\,\,\right] \,\, 
Z_{\rm CFT}\left[\,\,
\lower0.15in\hbox{\epsfig{figure=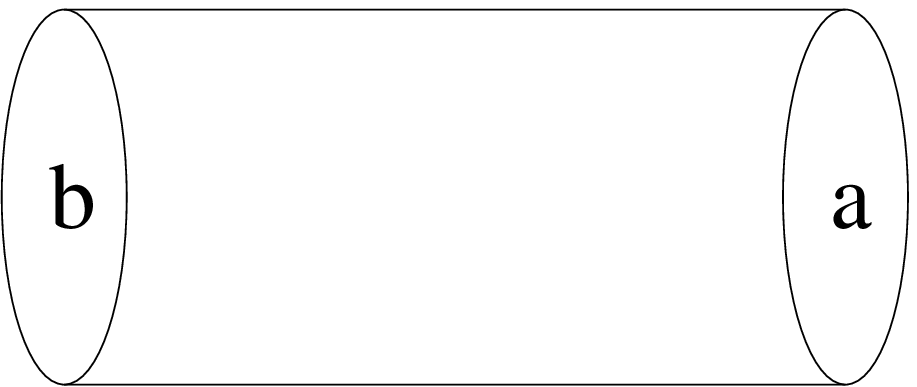,height=0.3in}}
\,\,\right]=
Z_{\rm CFT}\left[\,\,
\lower0.15in\hbox{\epsfig{figure=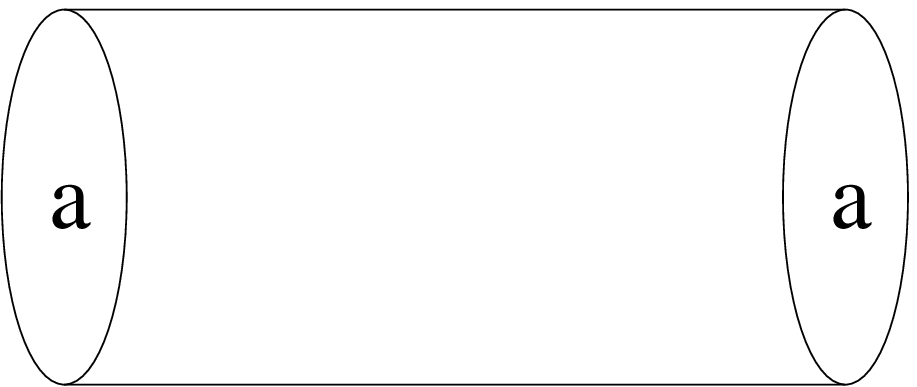,height=0.3in}}
\,\,\right]
\end{equation}
One applies a similar procedure to the state of the black hole with
point particle, to obtain the density matrix as the CFT partition
function on the ``long'' cylinder with two vertex operators inserted:
\begin{equation}
\rho_{BH-pp}=
Z_{\rm CFT}\left[\,\,
\lower0.15in\hbox{\epsfig{figure=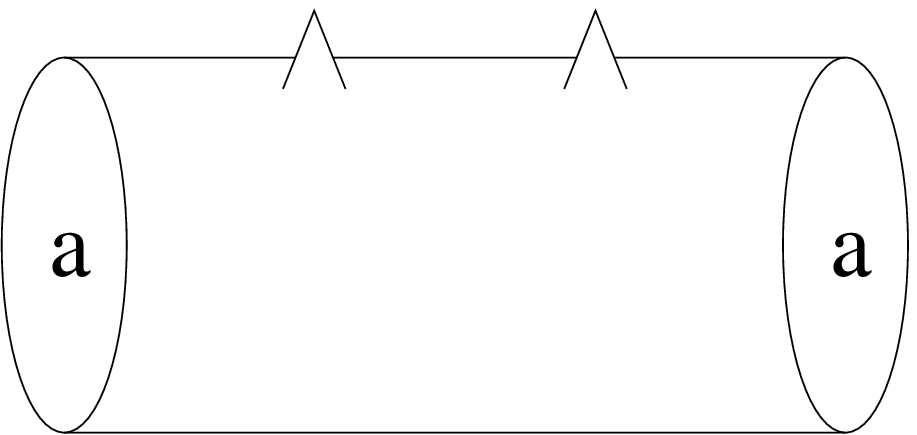,height=0.3in}}
\,\,\right]
\end{equation}
The probability of the emission process is now given by the overlap
of these two density matrices, divided by the traces of each of
these matrices:
\begin{equation}\label{emission-prob}
{\cal P}= \frac{{\rm Tr}\left( \rho_{BH} \, \rho_{BH-pp} \right)}
{{\rm Tr}(\rho_{BH}){\rm Tr}(\rho_{BH-pp})}.
\end{equation}
This is related to the quantity (5.2). The later is, however,
not correctly normalised to be interpreted as the probability.
The arguments given later in section 5.1 still apply and
show that (\ref{emission-prob}) is qualitatively correct.
\item The prescription given in section 5.2 should also be modified.
The expression 5.11 for the state of AdS with two point particles
in it is correct. However, to extract the black hole creation
probability one should follow a procedure similar to the one above.
We again introduce the density matrix of the black hole given
by (\ref{density-BH}). One can similarly create the density matrix
describing the pure state (5.11):
\begin{equation}
\rho_{2pp} = 
Z_{\rm CFT}\left[\,\,
\lower0.16in\hbox{\epsfig{figure=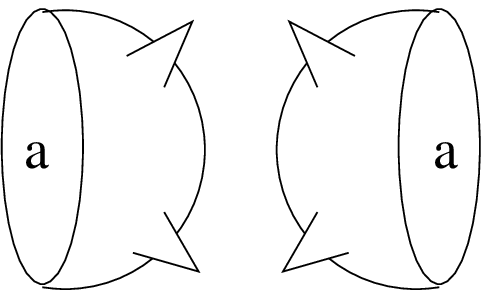,height=0.4in}}
\,\,\right]
\end{equation}
The probability of black hole creation is given by the overlap of the
density matrices of the black hole and that of the pair of point
particles, properly normalised:
\begin{equation}\label{creation-prob}
{\cal P}= \frac{{\rm Tr}\left( \rho_{BH} \, \rho_{2pp} \right)}
{{\rm Tr}(\rho_{BH}){\rm Tr}(\rho_{2pp})}.
\end{equation}
The quantity in the numerator is the 4-point function (5.12) projected
on a particular intermediate state (determined by the black hole size).
The quantities in the denominator are the CFT partition function
on the torus, and the partition function on the 4-punctured sphere
correspondingly. The arguments of the second paper can still be
applied to this expression, and conclusions will be analysed below.
\end{itemize}

\bigskip
Here are the corrections for the second paper.
\begin{itemize}
\item The prescription for computing the black hole creation
probability given in the introduction should be replaced
by the prescription (\ref{creation-prob}) above. The
prescription (1.3) is similar, but incorrectly normalised.
Thus, the derivation of the answer (1.4) should be modified.
The correct semi-classical regime answer can be obtained
very easily as follows. In the semi-classical limit
the 4-point function is peaked on a particular
intermediate state - the one that corresponds to 
the black hole that would be created in the process.
Therefore, in the expression for the probability of creating this black 
hole the numerator of (\ref{creation-prob}), which is the 4-point 
function projected on a particular intermediate state, is approximately
equal to the second term in the denominator, which is the full
4-point function. Thus, in the semi-classical regime, 
the creation probability is given by inverse of
the black hole partition function:
\begin{equation}\label{prob}
{\cal P}\sim 1/Z_{BH}=e^{I_{BH}}=e^{-S_{BH}/2}=e^{-\beta M}.
\end{equation}
Here $Z_{BH}, I_{BH}, S_{BH}$ are the BH partition function, free energy and
entropy correspondingly, and $\beta, M$ are the inverse temperature and mass
of the BH being created.
This last expression coincides with the one given by (1.4). Note, however,
that it is incorrectly stated in the text that the probability is given by
$e^{-\beta M/2}$. The correct answer is given by (\ref{prob}). The 
arguments of the remainder of the introduction go through almost unchanged.
\item The analysis of the rest of the paper is essentially correct, but
as is clear from the previous argument, is not necessary in the
semi-classical regime. The contents of the sections 2-3 are interesting in its own right,
as they deal with the classical theory. We believe that the contents of sections 5-6
are still of interest, as they demonstrate the problems that will have to
be overcome for a full quantum computation of the probability to
be performed. The analysis of this sections is also necessary to
confirm that the probability is peaked on the black hole that
would be created in the corresponding classical process. The analysis
of the probability in section 7 needs to be modified in that the normalising
factors (\ref{creation-prob}) should be introduced. For this reason,
the answer (7.7) needs to be modified, with the correct answer
given above by (\ref{prob}). Let us also note that there is a
sign mistake in section 7 getting (7.7) as the value of the 4-point
function. The logarithm of such a 4-point function for the case of maximally massive
particles is equal (in the semi-classical regime) to half of
the logarithm of the black hole partition function. Thus, the
semi-classical limit of the 4-point function for such particles
is given by an exponential with a {\it positive} large argument.
This answer makes it obvious that normalization factors such as
ones in (\ref{creation-prob}) are necessary to obtain the
probability.
\item The argument in the discussion section, page 4023, paragraph before last on this page, has to
be modified in view of the fact that the correct answer is given by
(\ref{prob}). Thus, the exponentially small probability is not surprising 
in view of the fact that the probability we got can
also be interpreted as the probability of a BH of mass $M$ blowing up into
two particles. The probability $\exp{-\beta M}$ we got is that of a thermal emission
of all of $BH$ mass $M$ at temperature $1/\beta$, which seems a natural answer.
\end{itemize}

\section*{Acknowledgements}

The author is supported by the EPSRC advanced fellowship.

\end{document}